\documentclass[aps,prb,twocolumn,superscriptaddress,showpacs]{revtex4}

\usepackage{graphicx, bm, amsmath}

\newcommand{\me}{\mathrm{e}}
\newcommand{\mi}{\mathrm{i}}

\newcommand{\dif}{\mathrm{d}}

\newcommand{\s}{\hspace{1ex}}

\begin{document}


\title{Cooperative Jahn-Teller Distortion in PrO$_2$}



\author{C.\,H. Gardiner}
\email{carol.webster@npl.co.uk}
\affiliation{National Physical Laboratory, Queens Road, Teddington, Middlesex, TW11 0LW, UK}

\author{A.\,T. Boothroyd}
\affiliation{Clarendon Laboratory, University of Oxford, Parks Road, Oxford, OX1 3PU, UK}

\author{P. Pattison}
\affiliation{Swiss-Norwegian Beamline, European Synchrotron Radiation Facility, Bo{\^i}te Postale 220, F-38043, Grenoble C{\'e}dex 9, France}

\author{M.\,J. McKelvy}
\affiliation{Center for Solid State Science, Arizona State University, Tempe, Arizona 85287-1704}

\author{G.\,J. McIntyre}
\affiliation{Institut Laue-Langevin, Bo{\^i}te Postale 156, F-38042, Grenoble C{\'e}dex 9, France}

\author{S.\,J.\,S. Lister}
\affiliation{Oxford Magnet Technology Ltd, Wharf Road, Eynsham, Witney, Oxfordshire, OX29 4BP, UK}



\date{\today}

\begin{abstract}
We report neutron diffraction data on single crystal PrO$_2$ which
reveal a cooperative Jahn-Teller distortion at $T_{\rm D} = 120
\pm 2$\,K.  Below this temperature an internal distortion of the
oxygen sublattice causes the unit cell of the crystallographic
structure to become doubled along one crystal axis.  We discuss
several possible models for this structure.  The antiferromagnetic
structure below $T_{\rm N} = 13.5$\,K is found to consist of two
components, one of which shares the same doubled unit cell as the
distorted crystallographic structure. We also present measurements
of the magnetic susceptibility, the specific heat capacity and the
electrical conductivity of PrO$_2$. The susceptibility data show
an anomaly at a temperature close to $T_{\rm D}$.  From the
specific heat capacity data we deduce that the ground state is
doubly degenerate, consistent with a distortion of the cubic local
symmetry.  We discuss possible mechanisms for this.  The
conductivity shows an activated behaviour with an activation
energy $E_{\rm a} = 0.262 \pm 0.003$\,eV.
\end{abstract}



\maketitle


\section{Introduction}

In recent years there has been a resurgence of interest in
phenomena associated with orbital degrees of freedom,
\cite{Tokura} such as orbital ordering, orbital waves
(``orbitons") \cite{Saitoh} and unusual spin-orbital liquid ground
states. \cite{Khaliullin:2000, Khaliullin:2001} Much of the work
has been stimulated by studies of the perovskite-based
transition-metal oxides, e.g. the colossal magnetoresistance
manganites, whose properties can be strongly influenced by the
extent to which orbital degrees of freedom are quenched by
Jahn-Teller distortions of the lattice or coupled to other
electronic degrees of freedom.

Jahn-Teller and orbital phenomena also occur in compounds
containing localized 4$f$ and 5$f$ electrons. Among the simplest
of these, the fluorite-structure actinide dioxides UO$_2$ and
NpO$_2$ have been investigated for many years, \cite{Santini,
Paixao} and are now understood to exhibit complex ordered phases
at low temperatures involving coupled electric and magnetic
multipoles as well as (in the case of UO$_2$) a lattice
distortion.

Unusual magnetic effects have also been found in the
fluorite-structure lanthanide dioxide PrO$_2$. Some years ago,
PrO$_2$ was found to have an anomalously small ordered moment in
the antiferromagnetic phase. \cite{Kern:1984}  Very recently, we
discovered a broad continuum in the magnetic excitation spectrum
of PrO$_2$ probed by neutron inelastic scattering, which we
ascribed to Jahn-Teller fluctuations involving the
orbitally-degenerate 4$f$ ground state and dynamic distortions of
the lattice. \cite{Boothroyd:2001} In a separate neutron
diffraction experiment \cite{Gardiner:2002:PrO2} we found evidence
that the antiferromagnetic structure contains a component with
twice the periodicity of the accepted type-I magnetic structure.
\cite{Kern:1984}

In this paper we follow up our previous work \cite{Boothroyd:2001,
Gardiner:2002:PrO2} with further experimental results on the
electronic and magnetic behaviour of PrO$_2$. Our main results
come from neutron and x-ray diffraction measurements of the
crystallographic and magnetic structure of PrO$_2$ in the
temperature range 2--300\,K. These reveal the existence of an
internal distortion of the fluorite structure below $T_{\rm D} =
120$\,K and a related distortion of the antiferromagnetic
structure below $T_{\rm N} = 13.5$\,K. We also report measurements
of the magnetic susceptibility, the specific heat capacity, and
the electrical conductivity of PrO$_2$, which allow us to
determine the effective paramagnetic moment of the Pr ion, the
degeneracy of the 4$f$ electron ground state, and the activation
energy of the charge carriers.

\section{Sample preparation}

Measurements were made on single crystal and polycrystalline
samples of PrO$_2$.  The single crystals used for the neutron
diffraction and conductivity measurements were taken from a batch
prepared several years ago by a hydrothermal procedure.
\cite{McKelvy} The sample used for neutron diffraction was the
largest of the batch, with a mass of $\approx 1$\,mg, while the
sample used for the conductivity measurements was somewhat
smaller.  Both were irregularly shaped.  Polycrystalline samples
were used for the x-ray diffraction, magnetic susceptibility and
specific heat capacity measurements.  These were prepared by
oxidation of commercially obtained Pr$_6$O$_{11}$. The starting
powder was first baked in air at 1000$^{\circ}$C for $\sim$ 11
hours, then annealed in flowing oxygen at 280$^{\circ}$C for 30
days.  The powder was reground approximately once a week during
the annealing.  Synchrotron x-ray diffraction showed that the
final product contained $\sim 1$\% of residual Pr$_6$O$_{11}$.

\section{Structural Investigation}

We investigated the crystallographic and magnetic structure of
PrO$_2$ by performing a neutron diffraction study on a single
crystal sample. The experiment was carried out on the D10
four-circle diffractometer at the Institut Laue-Langevin.  We used
an Eulerian cradle for crystal orientation and a position
sensitive detector.  The latter was kept in the scattering plane
at all times. The diffractometer configuration was as follows:
vertically curved pyrolytic graphite (002) or Cu (200)
monochromator, no collimators, circular aperture of diameter 6\,mm
before the sample and rectangular aperture of dimensions 20\,mm
$\times$ 25\,mm before the detector.  We used incident neutron
wavelengths of 2.3575\,\AA\ ($\lambda_1$) and 1.2579\,\AA\
($\lambda_2$) from the pyrolytic graphite and Cu monochromators
respectively, with a pyrolytic graphite filter for the former to
reduce the half-wavelength contribution. The larger wavelength
provided greater flux, while the smaller provided greater coverage
of reciprocal space.

The single crystal sample was mounted on a thin aluminium pin and
aligned such that the [1$\bar{1}$0] direction lay along the axis
of the pin.  The pin was attached to the Eulerian cradle inside a
helium flow cryostat.

\subsection{Measurements}

The quality of the crystal was checked by performing crystal
rotation scans ($\omega$-scans) of the strongest structural Bragg
reflections at room temperature.  The peak widths were found to be
within the experimental resolution.

To examine the crystallographic structure of PrO$_2$, as many
structural Bragg reflections as possible were measured by
$\omega$-scan at the accessible positions in reciprocal space
satisfying the selection rules for the fluorite structure ($h$,
$k$ and $l$ all even or all odd). This set of reflections was
measured at $\lambda_1$ at temperatures of 2\,K, 20\,K, 90\,K,
150\,K and 300\,K. At some temperatures the full set was measured
at both neutron wavelengths ($\lambda_1$ and $\lambda_2$).

\begin{figure}[!ht]
\begin{center}
\includegraphics{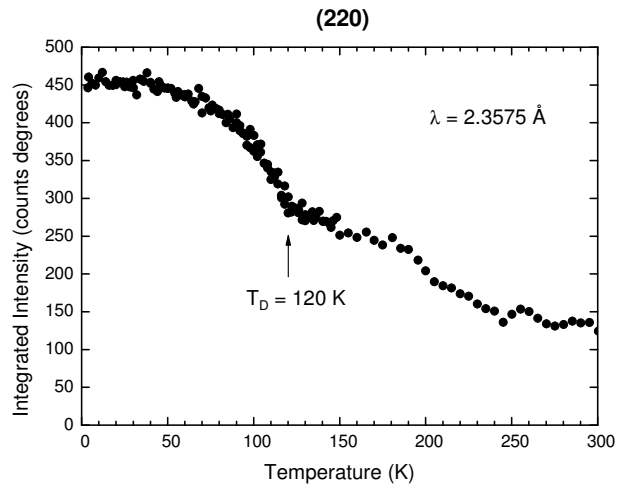}
\caption{Temperature dependence of the (220) reflection.}
\label{fig:(220)vsT}
\end{center}
\end{figure}

The intensities of the strongest reflections (e.g. (220), (400))
were found to change rapidly with temperature.  The temperature
dependence of these reflections was measured more carefully by
performing $\omega$-scans at 2\,K intervals between 2\,K and
130\,K, then at 5\,K intervals between 130\,K and 300\,K. Figure
\ref{fig:(220)vsT} shows the integrated intensity of the scan as a
function of temperature.  A sharp rise in intensity can be seen
below 120\,K.  We found that the rise was greatest for the
strongest reflections and more pronounced at the larger wavelength
$\lambda_1$, suggesting a sudden reduction in extinction.  A
possible explanation for this observation is a reduction in grain
size caused by the strain associated with a structural distortion.

\begin{figure}[!ht]
\begin{center}
\includegraphics{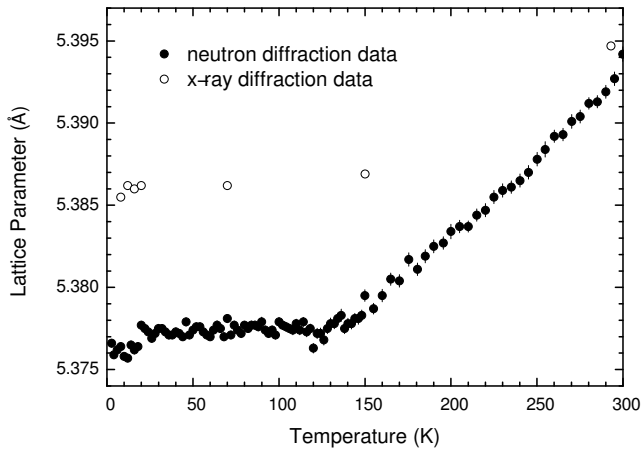}
\caption{Temperature dependence of the lattice parameter in
PrO$_2$.}
\label{fig:LatticeParvsT}
\end{center}
\end{figure}

The temperature dependence of the lattice parameter, estimated
from the variation of the Bragg angle with temperature for the
(400) reflection, also shows a sharp change at $\sim$ 120\,K (see
Fig.\ \ref{fig:LatticeParvsT}: closed circles).  Although the
absolute value of the lattice parameter obtained by this method is
inaccurate, as it relies on the crystal being perfectly aligned,
the data nevertheless give a good qualitative indication of the
way the lattice parameter changes with temperature.  It is seen to
decrease linearly from 300\,K to $\sim 120$\,K, below which it
becomes approximately constant, dipping slightly at $\sim 18$\,K.
The open circles in Figure \ref{fig:LatticeParvsT} show data taken
from a high-resolution x-ray powder diffraction study on PrO$_2$
performed at the Swiss-Norwegian Beamline BM1B at the European
Synchrotron Radiation Facility (ESRF).  There are fewer points in
this plot, but the absolute value of the lattice parameter is much
more accurate because it has been obtained by Rietveld refinement
of the diffraction pattern.  The temperature variation found by
x-ray diffraction is consistent with that found by neutron
diffraction.

To probe the magnetic structure, as many magnetic Bragg
reflections as possible were measured by $\omega$-scan at the
accessible positions in reciprocal space satisfying the selection
rules for the antiferromagnetic (AFM) type-I magnetic structure
($h$, $k$ and $l$ a mixture of even and odd). This set of
reflections was measured at temperatures of 2\,K and 20\,K (below
and above $T_{\rm N}$) using only the larger wavelength
$\lambda_1$, since this provided higher flux.

In addition to the above measurements we also performed
$\omega$-scans at a number of positions in reciprocal space with
half-integer Miller indices.  The object of these measurements was
to search for peaks seen previously in a powder diffraction
experiment. \cite{Gardiner:2002:PrO2}  The scans were performed at
temperatures of 2\,K, 20\,K and 150\,K.

\begin{figure}[!ht]
\begin{center}
\includegraphics{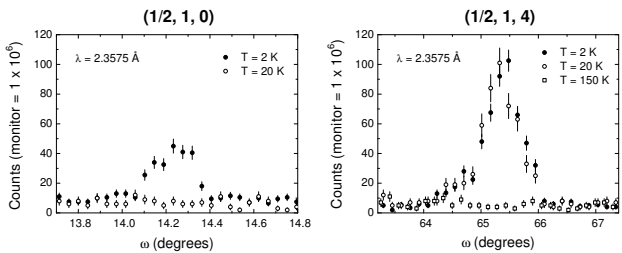}
\caption{$\omega$-scans of typical half-integer reflections
(detector counts are normalised to a fixed incident beam monitor
count of $1 \times 10^6$, corresponding to a counting time of
$\sim$ 100 seconds). Those reflections with $l=0$ are absent above
$T_{\rm N}$. Those with $l \ne 0$ are still present above $T_{\rm
N}$. All are absent at $T$ = 150\,K.}
\label{fig:HalfIntPeaks2and20K}
\end{center}
\end{figure}

At $T$ = 2\,K, we found reflections at positions satisfying the
selection rule $h=n+1/2$, $k$ = odd, $l$ = even, where $n$, $k$
and $l$ are integers, and $h$, $k$ and $l$ can be commuted.  At $T
= 20$\,K (above $T_{\rm N}$), we found reflections at exactly the
same positions, except that those with $l=0$ were no longer
present.  The strongest half-integer reflection we observed was
$\sim$1\% of the intensity of the strongest fluorite reflection.
At $T = 150$\,K, all the half-integer reflections had disappeared.
Figure \ref{fig:HalfIntPeaks2and20K} shows $\omega$-scans of the
$\left(\frac{1}{2}10\right)$ and $\left(\frac{1}{2}14\right)$
reflections.  These are typical of the two observed sets of
half-integer reflections: those with $l=0$ and those with $l \ne
0$.

We measured the temperature dependence of the
$\left(\frac{1}{2}10\right)$ reflection by counting at the peak
centre at a series of temperatures between 2\,K and 20\,K. The
temperature dependence of the stronger
$\left(\frac{1}{2}14\right)$ reflection was measured by performing
$\omega$-scans at a series of temperatures between 2\,K and
150\,K.

\begin{figure}[!ht]
\begin{center}
\includegraphics{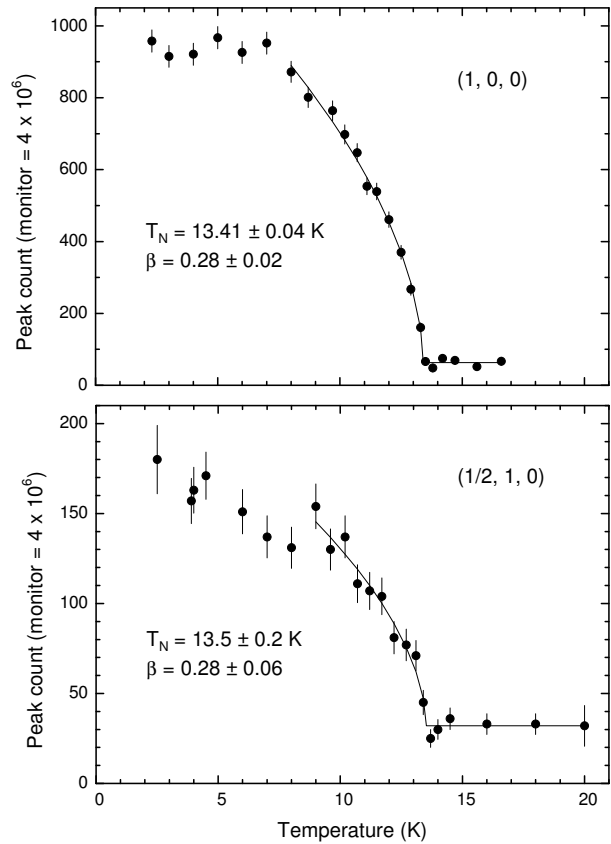}
\caption{Temperature dependence of the $\left(\frac{1}{2} 1
0\right)$ and (100) intensities (detector counts are normalised to
a fixed incident beam monitor count of $4 \times 10^6$,
corresponding to a counting time of $\sim 8$ minutes in the upper
graph and $\sim 6.5$ minutes in the lower graph).  Order parameter
curves of the form $I \propto (T_{\rm N} - T)^{2\beta}$ have been
fitted to the data to determine $T_{\rm N}$.  Note that the
temperature dependence of the (100) reflection was actually
measured during an earlier experiment performed on the same
instrument with the same crystal, but with a different sample
environment, so the intensities in the two plots should not be
compared directly.}
\label{fig:(1-210)and(100)OrderPar}
\end{center}
\end{figure}

Figure \ref{fig:(1-210)and(100)OrderPar} compares the temperature
dependence of the $\left(\frac{1}{2}10\right)$ reflection with
that of the magnetic reflection (100).  To within experimental
uncertainty, both reflections disappear at the antiferromagnetic
ordering temperature $T_{\rm N}$.  We were able to determine
$T_{\rm N}$ for both reflections by fitting the data using order
parameter curves of the form

\begin{equation}
I \propto (T_{\rm N} - T)^{2\beta},
\end{equation}

\noindent where $I$ is the peak count and $\beta$ is the critical
exponent.  For the (100) reflection we found $T_{\rm N} = 13.41
\pm 0.04$\,K, while for the $\left(\frac{1}{2} 1 0\right)$
reflection we found $T_{\rm N} = 13.5 \pm 0.2$\,K.  Since these
are in excellent agreement, it is likely that both reflections are
magnetic in origin and belong to the same magnetic phase.  The
existence of half-integer magnetic Bragg reflections indicates
that the magnetic unit cell is doubled along one crystal axis.

\begin{figure}[!ht]
\begin{center}
\includegraphics{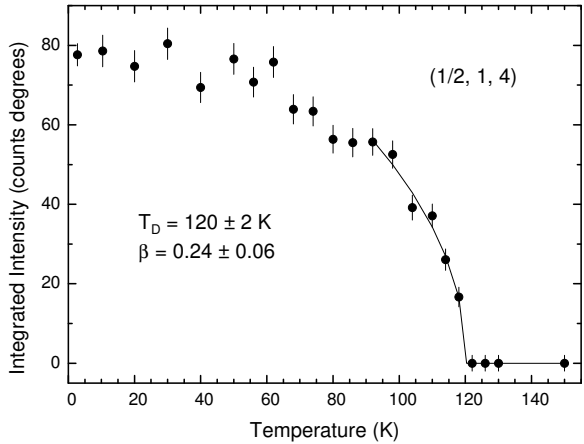}
\caption{Temperature dependence of the $\left(\frac{1}{2} 1
4\right)$ intensity.  An order parameter curve of the form $I
\propto (T_{\rm D} - T)^{2\beta}$ has been fitted to the data to
determine $T_{\rm D}$.}
\label{fig:(1-214)PeakAndOrderPar}
\end{center}
\end{figure}

Figure \ref{fig:(1-214)PeakAndOrderPar} shows the
temperature-dependence of the $\left(\frac{1}{2} 1 4\right)$
reflection.  It is found to disappear at $T_{\rm D} = 120 \pm
2$\,K, which is the same as the temperature below which we
observed a sharp rise in the intensities of the large structural
reflections.  This strongly suggests that the $\left(\frac{1}{2} 1
4\right)$ reflection is structural in origin.  The existence of
half-integer structural reflections indicates that the
crystallographic unit cell is also doubled along one crystal axis.

A re-examination of previous data taken with a powder sample on
the POLARIS diffractometer at the ISIS facility
\cite{Gardiner:2002:PrO2} has now also revealed half-integer
structural reflections.  Although two of the magnetic half-integer
reflections had been seen previously, the structural reflections
had not been noticed, due to the nearby presence of other peaks of
similar intensity, which were probably due to scattering from the
sample environment. The x-ray powder diffraction study mentioned
above also confirms the presence of half-integer structural
reflections.

\subsection{Structure analysis}

\label{sec:analysis}

To identify the nature of the crystallographic structure below
$T_{\rm D} = 120$\,K and the magnetic structure below $T_{\rm N} =
13.5$\,K, we performed a detailed analysis of the intensities of
the observed integer and half-integer structural and magnetic
reflections at low temperatures.

After background subtraction, each $\omega$-scan was integrated
using the extended trapezoidal rule \cite{NumericalRecipes} (a
simple method of numerical integration), corrected for the Lorentz
factor, normalised to the monitor count and the intensity of the
(311) reflection, and multiplied by a wavelength-dependent scale
factor to enable direct comparison between the intensities of the
observed structural reflections and their calculated structure
factors. The (311) reflection was used for normalisation because
it was one of the weakest reflections accessible at both
wavelengths, so it was little affected by extinction.  In
addition, its structure factor contained no contribution from the
oxygen ions, so it was unaffected by the distortion.  Finally, the
corrected and scaled intensities were averaged over
symmetry-equivalent reflections to guard against the possibility
that a chance uneven distribution in symmetry-equivalent
structural or magnetic domain populations might affect the
measured intensity of any given reflection (although this was
unlikely since the majority of the symmetry-equivalent reflections
were equal in intensity to within experimental uncertainty).

We then considered distortions of the fluorite crystallographic
structure and the type-I antiferromagnetic structure which would
give rise to structure factors in agreement with the relative
intensities and selection rules of the observed reflections.  Such
structures were identified by trial and error.  However, in the
case of the crystallographic structure we were guided by the
results of the x-ray diffraction study, which revealed no
splitting of the reflections caused by the fluorite structure.
This indicated that the Pr lattice remained undisturbed, and any
distortion must therefore be due to an internal rearrangement of
the oxygen atoms.

Once a suitable distorted crystallographic structure was found, we
refined the overall magnitude of the oxygen displacements by a
least squares method to achieve the best possible agreement
between the structure factors and peak intensities.  For the
magnetic structure we used the ratio of the peak intensity to the
magnetic structure factor for each reflection to obtain a value
for the ordered moment $\mu$ of the Pr ion.

The structure factors of the structural reflections are given by

\begin{equation}
|F_{\rm N}({\bf Q})|^2 = \biggl|\sum_{j}\left\langle\bar{b}_j
\me^{\mi{\bf Q}.{\bf r}_j}\me^{-W_j(Q,T)}\right\rangle\biggr|^2\,
\label{eq:nucstrucfac}
\end{equation}

\noindent where the summation index $j$ runs over all the atoms in
the unit cell, $\bar{b}_j$ is the nuclear scattering length of the
$j$th atom averaged over all of its isotopes, ${\bf r}_j$ is the
position of the $j$th atom within the unit cell, and
$\me^{-W_j(Q,T)}$ is the Debye-Waller factor (we set this equal to
1, since the measurements were made at low temperatures) and
$\langle \rangle$ denotes an average over all symmetry-equivalent
structural domains.  The scaled, integrated intensities of the
structural reflections are equal to $|F_{\rm N}({\bf Q})|^2$. The
structure factors of the magnetic reflections are given by

\begin{equation}
|F_{\rm M}({\bf Q})|^2 =
\sum_{\alpha\beta}\left\langle\left(\delta_{\alpha\beta} -
\hat{Q}_{\alpha}\hat{Q}_{\beta}\right)F_{\rm M}^{\alpha}({\bf
Q})F_{\rm M}^{\beta*}({\bf Q})\right\rangle,
\label{eq:average}
\end{equation}

\noindent where the summation indices $\alpha$ and $\beta$ run
over the cartesian co-ordinates $x,y$ and $z$,
$\delta_{\alpha\beta}$ is the Kronecker delta, * denotes the
complex conjugate, $\hat{Q}_{\alpha}$ is the $\alpha$-component of
the unit scattering vector and $\langle \rangle$ denotes an
average over all symmetry-equivalent magnetic domains.  In the
dipole approximation $F_{\rm M}^{\alpha}({\bf Q})$ is given by

\begin{equation}
F_{\rm M}^{\alpha}({\bf Q}) = f({\bf
Q})\sum_{j}\hat\mu_j^{\alpha}\me^{\mi{\bf Q}.{\bf
r}_j}\me^{-W_j(Q,T)},
\end{equation}

\noindent where the summation index $j$ runs over all the magnetic
atoms in the magnetic unit cell, $\hat\mu_j^{\alpha}$ is the
$\alpha$-component of a unit vector in the direction of the
magnetic moment of the $j$th magnetic atom, ${\bf r}_j$ is the
position of the $j$th magnetic atom within the magnetic unit cell,
$f({\bf Q})$ is the magnetic form factor of the Pr$^{4+}$ ion and
$\me^{-W_j(Q,T)}$ is the Debye-Waller factor (set equal to 1). The
scaled, integrated intensities of the magnetic reflections are
proportional to $\big(\frac{\gamma
r_0}{2}\big)^2\big(\frac{\mu}{\mu_{\rm B}}\big)^2|F_{\rm M}({\bf
Q})|^2$, where $\big(\frac{\gamma r_0}{2}\big)^2 = 72.4$\,mb ($=
7.24$\,fm$^2$) and $\mu_{\rm B}$ is the Bohr magneton.

For reference we note the structure factors of the fluorite
structure here:

\begin{equation}
|F_{\rm N}({\bf Q})|^2 = \begin{cases} \s \text{336\,fm}^2, &
\text{$h$, $k$, $l$ all odd}\\ \s \text{791\,fm}^2, & h+k+l = 2n\\
\text{4194\,fm}^2, & h+k+l = 4n.\end{cases}
\end{equation}

\noindent These were computed using the fluorite unit cell, which
contains four Pr ions and eight O ions.  For the half-integer
reflections (both structural and magnetic) we used a unit cell
that was doubled along one axis, thus containing eight Pr ions and
sixteen O ions.  In order to allow direct comparison with the
structure factors of the fluorite structure we divided the
structure factors $F_{\rm N}({\bf Q})$ and $F_{\rm M}({\bf Q})$ of
the half-integer reflections, calculated in the doubled cell, by
2.

\subsubsection{Distorted crystallographic structure}

\begin{figure}[!ht]
\begin{center}
\includegraphics{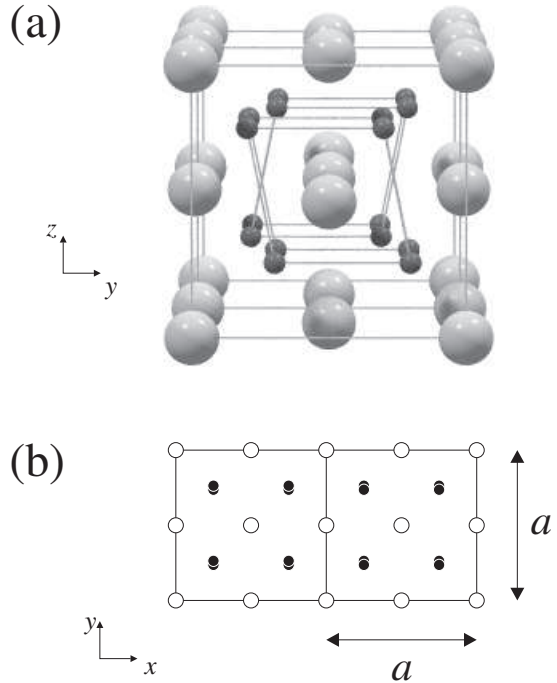}
\caption{Sheared structure. (a) The large pale spheres are
praseodymium ions and the small dark spheres are oxygen ions. The
oxygen cubes in the two halves of the doubled unit cell are
sheared in opposite senses, but both have a shearing vector that
is perpendicular to the direction along which the cell is doubled.
The displacements of the oxygen ions have been exaggerated for
clarity. (b) Scale diagram showing the distorted structure in plan
view.  The white circles are praseodymium ions and the black
circles are oxygen ions.}
\label{fig:distortedstructures}
\end{center}
\end{figure}

One of the simplest structures that possesses structure factors in
agreement with the intensities of the half-integer structural
reflections is shown in Fig.\ \ref{fig:distortedstructures}(a).
The Pr lattice is undisturbed, but the oxygen ions are each
displaced by an amount $d$ in a direction perpendicular to that
along which the unit cell is doubled (in Fig.\
\ref{fig:distortedstructures} the doubling is along the
$x$-direction and the displacements are along the $y$-direction).
The oxygen cubes in the two halves of the doubled unit cell are
sheared in opposite senses, so we will refer to this structure as
the ``sheared" structure.  The space group is Imcb (a variant of
Ibam, space group 72), with the origin shifted by $\frac{1}{4}$ of
the unit cell length along the orthorhombic $a$-direction (the
$x$-direction in Fig.\ \ref{fig:distortedstructures}).

A simple least squares refinement of the oxygen displacement $d$,
based on comparison of the domain-averaged structure
factors\footnote{Averaging over symmetry-equivalent domains in
Equation \eqref{eq:nucstrucfac} means that we compare the observed
intensity $I(hkl)_{\rm obs}$ with the calculated intensity
$\frac{1}{6}[I(hkl)_{\rm calc} + I(klh)_{\rm calc} + I(lhk)_{\rm
calc} + I(\bar{h}lk)_{\rm calc} + I(\bar{k}hl)_{\rm calc} +
I(\bar{l}kh)_{\rm calc}]$, where $h$, $k$ and $l$ are in the cubic
cell.  This accounts for three-fold twinning around the cubic
$\langle 111\rangle$ direction and reflection across the cubic
$(01\bar{1})$ plane.} and the observed intensities of the
half-integer reflections, yielded $d = 0.0726(54)$\,\AA.  The
structure factors and observed intensities at $T = 20$\,K are
listed in Table \ref{tab:Half-intComparison}, and a plan view of
the structure with displacements to scale is shown in Fig.\
\ref{fig:distortedstructures}(b).  A full crystal structure
refinement using the Cambridge Crystallography Subroutine Library
(CCSL) yielded $d = 0.0664(17)$\,\AA.

\begin{table}[!ht]
\renewcommand{\arraystretch}{1.5}
\begin{center}
\begin{tabular}{|c||c|c|}
\hline
Reflection & $\left|\frac{1}{2}F_{\rm N}({\bf Q})\right|^2$ (fm$^2$) & Intensity (fm$^2$)\\
\hline \hline
$\left(\frac{1}{2}10\right)$ & \s 0.0 & \s 0.1 $\pm$ 0.1\\
\hline
$\left(\frac{3}{2}10\right)$ & \s 0.0 & \s 0.3 $\pm$ 0.1\\
\hline
$\left(\frac{5}{2}10\right)$ & \s 0.0 & \s 0.5 $\pm$ 0.2\\
\hline
$\left(\frac{1}{2}12\right)$ & \s 5.1 & \s 5.5 $\pm$ 0.3\\
\hline
$\left(\frac{3}{2}12\right)$ & \s 5.1 & \s 5.0 $\pm$ 0.3\\
\hline
$\left(\frac{5}{2}12\right)$ & \s 5.1 & \s 4.8 $\pm$ 0.5\\
\hline
$\left(\frac{7}{2}12\right)$ & \s 5.1 & \s 4.8 $\pm$ 0.6\\
\hline
$\left(\frac{1}{2}14\right)$ & 19.9 & 16.6 $\pm$ 0.9\\
\hline
$\left(\frac{3}{2}14\right)$ & 19.9 & 20.5 $\pm$ 1.2\\
\hline
$\left(\frac{1}{2}16\right)$ & 42.7 & 42.3 $\pm$ 2.3\\
\hline
$\left(\frac{1}{2}30\right)$ & \s 0.0 & \s 0.4 $\pm$ 0.2\\
\hline
$\left(\frac{1}{2}32\right)$ & \s 5.1 & \s 5.2 $\pm$ 0.6\\
\hline
$\left(\frac{3}{2}32\right)$ & \s 5.1 & \s 7.1 $\pm$ 0.6\\
\hline
$\left(\frac{5}{2}32\right)$ & \s 5.1 & \s 7.1 $\pm$ 1.0\\
\hline
$\left(\frac{7}{2}32\right)$ & \s 5.1 & \s 6.3 $\pm$ 1.1\\
\hline
$\left(\frac{1}{2}34\right)$ & 19.9 & 17.7 $\pm$ 1.5\\
\hline
$\left(\frac{3}{2}34\right)$ & 19.9 & 22.0 $\pm$ 1.5\\
\hline
$\left(\frac{3}{2}36\right)$ & 42.7 & 45.6 $\pm$ 3.0\\
\hline
$\left(\frac{1}{2}52\right)$ & \s 5.1 & \s 6.9 $\pm$ 1.2\\
\hline
$\left(\frac{5}{2}52\right)$ & \s 5.1 & \s 8.7 $\pm$ 1.6\\
\hline
$\left(\frac{7}{2}52\right)$ & \s 5.1 & \s 4.9 $\pm$ 1.8\\
\hline
$\left(\frac{1}{2}72\right)$ & \s 5.1 & 12.4 $\pm$ 2.2\\
\hline
\end{tabular}
\caption{Comparison between $\left|\frac{1}{2}F_{\rm N}({\bf
Q})\right|^2$ for the sheared structure and the intensities of the
observed half-integer structural reflections at $T = 20$\,K. The
$F_{\rm N}({\bf Q})$ have been divided by 2, as described in Sec.
\ref{sec:analysis}, to allow direct comparison between the two
columns.}
\label{tab:Half-intComparison}
\end{center}
\end{table}

An identical shearing of the oxygen cube was originally proposed
for an internal distortion observed in UO$_2$, \cite{Faber}
although in that compound there was no evidence for a doubling of
the unit cell.  In UO$_2$ the oxygen ions are displaced by
0.014\,\AA\ from their fluorite structure sites, so the oxygen
displacements in PrO$_2$ are five times as large, and this is
reflected in the high value of the transition temperature.  It is
now accepted that the oxygen configuration in UO$_2$ forms a
triple-${\bf q}$ structure. \cite{Burlet}  However, there is
currently no reason to assume that the same is true in PrO$_2$.

\begin{figure}[!ht]
\begin{center}
\includegraphics{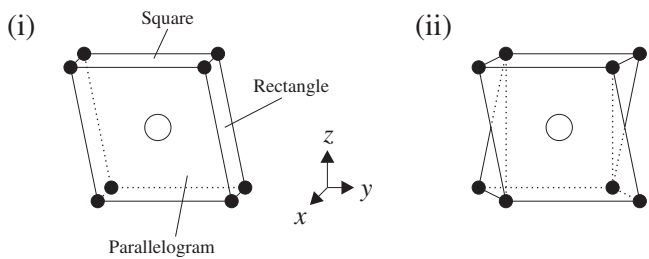}
\caption{The two different Pr sites that occur in the sheared
structure shown in Fig.\ \ref{fig:distortedstructures}.  The white
circles are praseodymium ions and the black circles are oxygen
ions. (i) Parallelohedron. (ii) Polyhedron with four parallelogram
faces and four triangular faces.}
\label{fig:PrSites}
\end{center}
\end{figure}

It is interesting to note that there are two distinct Pr sites in
the sheared structure, each of which occurs with equal frequency.
The two sites have different surrounding oxygen configurations,
which are shown in Fig.\ \ref{fig:PrSites}. At one site the
oxygens form a parallelohedron (a polyhedron with two square
faces, two rectangluar faces and two parallelogram faces), while
at the other they form a polyhedron with four parallelogram faces
and four triangular faces.

Although the distorted structure described above is consistent
with the intensities of the observed reflections, it is not the
only such structure.  For example, a superposition of two
structures identical to the sheared structure, but with oxygen
displacements of 0.0511\,\AA\ in mutually perpendicular
directions, gives similar agreement between the calculated
structure factors and observed intensities.  The overall
displacement of each oxygen ion in that case is $\sqrt{2}\times
0.0511 = 0.0722$\,\AA.

\begin{figure}[!ht]
\begin{center}
\includegraphics{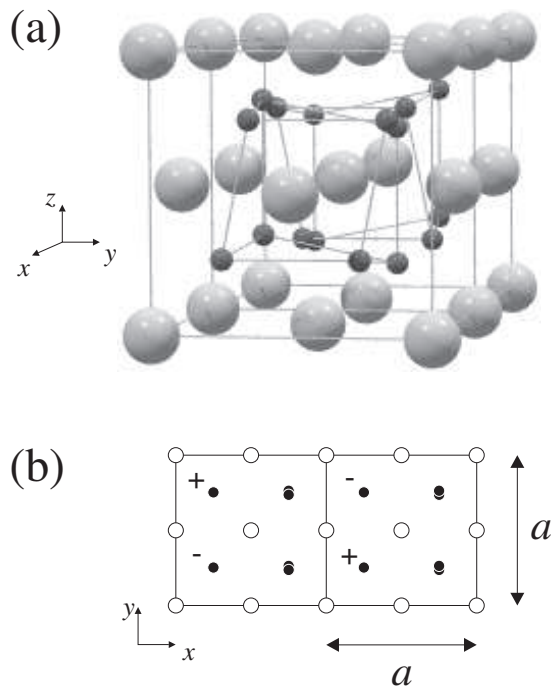}
\caption{Chiral distorted structure. (a) The large pale spheres
are praseodymium ions and the small dark spheres are oxygen ions.
The displacements of the oxygen ions have been exaggerated for
clarity. (b) Scale diagram showing the chiral structure in plan
view. The white circles are praseodymium ions and the black
circles are oxygen ions.  The $+$ and $-$ symbols indicate
positive and negative displacements along the $z$-direction.}
\label{fig:chiralstructure}
\end{center}
\end{figure}

The ``chiral" structure shown in Fig.\ \ref{fig:chiralstructure}
is also consistent with the intensities of the observed
reflections, and has the additional advantages of having a lower
Jahn-Teller energy \cite{Jensen} and a single Pr site.  A chiral
model is adopted for the oxygen displacement vector, but the
structure factors and the magnitude of the oxygen displacement are
almost identical to that of the sheared structure (a simple least
squares refinement yielded $d = 0.0726(54)$\,\AA, whereas a full
crystal structure refinement using CCSL yielded $d =
0.0663(16)$\,\AA). The space group is I4(1)/acd (space group 142),
with Pr at positions $8a$ $\left(0, \frac{1}{4},
\frac{3}{8}\right)$ and O at positions $16e$ $\left(\frac{1}{4}+d,
0, \frac{1}{4}\right)$, where the origin is chosen to be at the
centre of inversion, and $c$ is the unique axis.

\subsubsection{Doubled magnetic structure}

The presence of half-integer magnetic reflections alongside
stronger integer magnetic reflections indicates that the magnetic
structure consists of two components: a primary component with the
same unit cell as the fluorite structure and a secondary component
with a unit cell that is doubled along one crystal axis. This
picture fits well with the model of the distorted crystallographic
structure, in which the Pr lattice remains unchanged, but the
oxygen sublattice undergoes an internal distortion which gives
rise to a component of the structure with a doubled unit cell.

The easiest way to analyse the magnetic structure is to consider
the two components separately, so that the overall structure can
be visualised by performing a vector addition of the two
components of the magnetic moment for each Pr ion in the doubled
unit cell.

\begin{figure}[!ht]
\begin{center}
\includegraphics{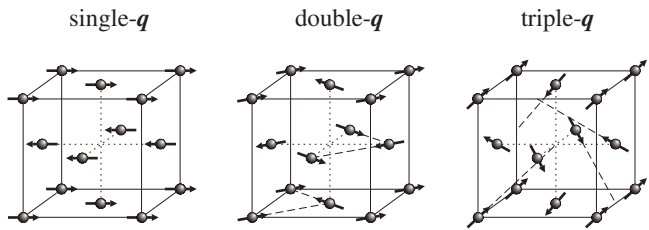}
\caption{Type-I primary component of the magnetic structure.
Neutron diffraction measurements under ambient conditions cannot
distinguish between the multi-${\bf q}$ structures shown, because
each gives rise to identical magnetic structure factors when
averaged over symmetry-equivalent magnetic domains. However, since
the crystallographic unit cell is doubled along one crystal axis,
the single-${\bf q}$ structure seems the most likely candidate.}
\label{fig:type-I}
\end{center}
\end{figure}

The primary component of the magnetic structure is the AFM type-I
structure, shown in Fig. \ref{fig:type-I}.  This gives rise to the
integer magnetic reflections.  Under ambient conditions it is not
possible to distinguish between the three multi-${\bf q}$
structures shown in the figure, but in view of the fact that the
structural distortion doubles the unit cell along one crystal axis
only, the single-${\bf q}$ structure seems the most likely
candidate.

The configuration of the magnetic moments in the secondary
component of the magnetic structure was deduced using a technique
similar to that for the structural distortion.  In our search for
possible configurations we found it impossible to conceive of a
magnetic structure that would give rise to reflections at $l=0$
alone. Symmetry requires that if a structure gives rise to
reflections at $l=0$ it will also give rise to reflections at all
positions with $l =$ even.  This means that many of the magnetic
half-integer reflections coincide with the structural ones.
However, since little change was observed in the intensities of
the structural half-integer reflections below $T_{\rm N}$ (see
Fig.\ \ref{fig:HalfIntPeaks2and20K}), it can be assumed that the
magnetic intensities at these positions are much smaller than the
structural intensities.  We have therefore made no attempt to
analyse the magnetic contribution at these positions. Our analysis
of the secondary component of the magnetic structure is based
solely on the half-integer reflections observed at $l=0$.

\begin{figure}[!ht]
\begin{center}
\includegraphics{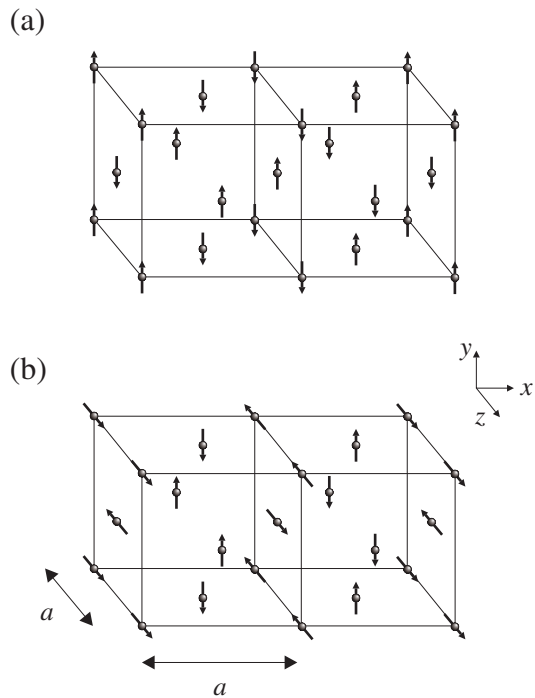}
\caption{Possibilities for the secondary component of the magnetic
structure.  (a) All moments point along the $y$-axis.  This
structure is associated with the chiral model for the structural
distortion shown in Fig.\ \ref{fig:chiralstructure}. (b) Half the
moments point along the $y$-axis and half along the $z$-axis. This
structure is associated with the model shown in Fig.\
\ref{fig:distortedstructures}.  The magnetic structure factors of
both structures are identical after domain averaging.}
\label{fig:DoubledMagStructures}
\end{center}
\end{figure}

The number of possible magnetic structures consistent with the
observed half-integer intensities is quite large.  Two of the
simplest possible structures are shown in Fig.\
\ref{fig:DoubledMagStructures}. Both possess identical magnetic
structure factors, shown in Table \ref{tab:DoubledMagStructures}.
Figure \ref{fig:DoubledMagStructures}(a) corresponds to the chiral
model for the structural distortion shown in Fig.\
\ref{fig:chiralstructure}, whereas Fig.\
\ref{fig:DoubledMagStructures}(b) corresponds to the model shown
in Fig.\ \ref{fig:distortedstructures}. The moments of the Pr ions
in both magnetic structures point along directions perpendicular
to the direction along which the unit cell is doubled.  Similar
structures in which some or all of the moments point along the
doubling direction give poorer agreement with the relative
intensities of the observed reflections.  It should be mentioned
that the $\left(\frac{5}{2} 1 0\right)$ reflection had a high and
rather sloping background, due to its proximity in reciprocal
space to an aluminium powder line, so its intensity is somewhat
unreliable.

\begin{table}[!ht]
\renewcommand{\arraystretch}{1.5}
\begin{center}
\begin{tabular}{|c||c|c|}
\hline
Reflection & $\left|\frac{1}{2}F_{\rm M}({\bf Q})\right|^2$ & Intensity (fm$^2$)\\
\hline \hline
$\left(\frac{1}{2}10\right)$ & 1.5 & 1.4 $\pm$ 0.1\\
\hline
$\left(\frac{3}{2}10\right)$ & 1.9 & 1.7 $\pm$ 0.1\\
\hline
$\left(\frac{5}{2}10\right)$ & 1.8 & 0.9 $\pm$ 0.3\\
\hline
$\left(\frac{1}{2}30\right)$ & 0.9 & 1.1 $\pm$ 0.2\\
\hline
$\left(\frac{3}{2}30\right)$ & 1.0 & 1.1 $\pm$ 0.2\\
\hline
\end{tabular}
\caption{For the doubled component of the magnetic structure the
magnetic structure factors $\left|\frac{1}{2}F_{\rm M}({\bf
Q})\right|^2$ (dimensionless) are compared with the magnetic
intensities measured at $T = 2$\,K. The $F_{\rm M}({\bf Q})$ have
been divided by 2, as described in Sec. \ref{sec:analysis} to
maintain consistency and with the structural calculations. The
ratio of magnetic intensity to magnetic structure factor is
constant, and can be used to determine the magnetic moment of the
Pr ion as described below.} \label{tab:DoubledMagStructures}
\end{center}
\end{table}

The magnetic moment of the Pr ion can be calculated for each of
the two components of the magnetic structure as follows:

\begin{equation}
\frac{\mu}{\mu_{\rm B}} = \sqrt{\frac{I_{\rm M}({\bf Q})|F_{\rm
N}({\bf Q}^{\prime})|^2}{\left(\frac{\gamma r_0}{2}\right)^2
I_{\rm N}({\bf Q}^{\prime})|F_{\rm M}({\bf Q})|^2}},
\end{equation}

\noindent where $\mu$ is the magnetic moment of the Pr ion in one
component of the magnetic structure, $I_{\rm M}({\bf Q})$ is the
integrated intensity of a magnetic reflection with reciprocal
lattice vector ${\bf Q}$, $|F_{\rm M}({\bf Q})|^2$ is the magnetic
structure factor of this reflection, $I_{\rm N}({\bf Q}^{\prime})$
is the integrated intensity of a nuclear reflection with
reciprocal lattice vector ${\bf Q}^{\prime}$ and $|F_{\rm N}({\bf
Q}^{\prime})|^2$ is the nuclear structure factor of this
reflection.

For each of the magnetic reflections we calculated the magnetic
moment of the Pr ion from the observed magnetic intensity and the
intensity of the (311) nuclear reflection.  The resulting values
of the Pr moment were averaged to reduce the uncertainty. For the
primary (AFM type-I) component of the magnetic structure we
obtained $\mu_1 = (0.65 \pm 0.02)\mu_{\rm B}$.  This is in good
agreement with a previous value of $(0.68 \pm 0.07)\mu_{\rm B}$
\cite{Gardiner:2002:PrO2} obtained by neutron diffraction from the
same single crystal sample. Previous powder diffraction
experiments have yielded values of $(0.6 \pm 0.1)\mu_{\rm B}$
\cite{Kern:1984} and $(0.572 \pm 0.012)\mu_{\rm B}$,
\cite{Gardiner:2002:PrO2} but it is likely that these are in
error, since in both cases the Rietveld refinement assumed the
crystallographic structure to be fluorite at low temperatures, and
therefore took no account of the structural distortion. For the
secondary (doubled) component of the magnetic structure we
obtained $\mu_2 = (0.35 \pm 0.04)\mu_{\rm B}$ (we disregarded the
unreliable $\left(\frac{5}{2} 1 0\right)$ reflection when
averaging the values of $\mu_2$ obtained from the half-integer
reflections).  The uncertainties on our values of $\mu_1$ and
$\mu_2$ are dominated by the uncertainty on the magnetic form
factor of Pr$^{4+}$.

\section{Bulk properties}

We now present measurements of some of the bulk properties of
PrO$_2$: magnetic susceptibility, specific heat capacity and
electrical conductivity.  These provide support for the findings
of the neutron diffraction experiments described above, and reveal
clues as to the origin of the structural distortion.

\subsection{Magnetic susceptibility}

\begin{figure}[!ht]
\begin{center}
\includegraphics{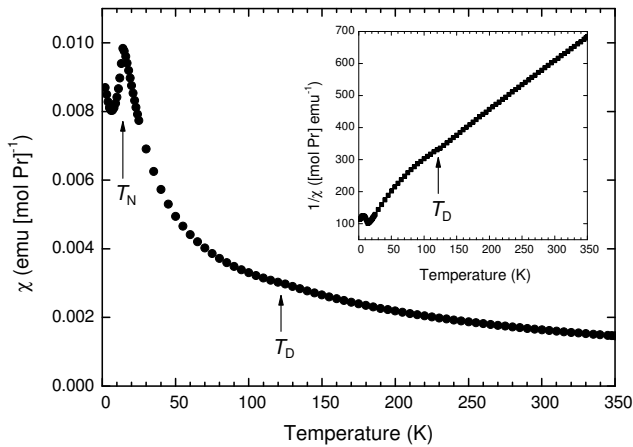}
\caption{Magnetic susceptibility of PrO$_2$ powder.  Molar
susceptibility is plotted in the main graph and inverse molar
susceptibility is shown in the inset.}
\label{fig:PrO2PowderSus}
\end{center}
\end{figure}

\begin{figure}[!ht]
\begin{center}
\includegraphics{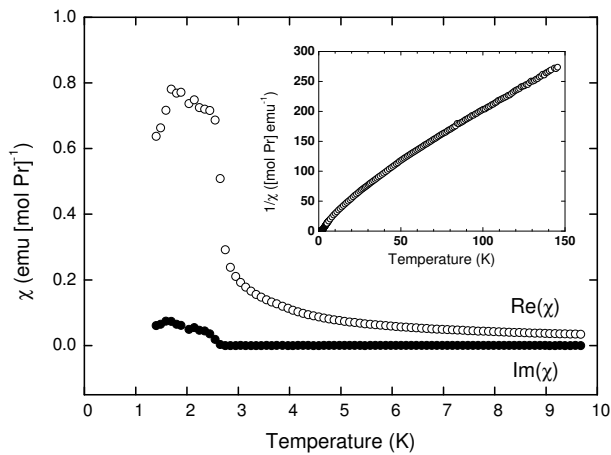}
\caption{AC magnetic susceptibility of Pr$_6$O$_{11}$ powder.  The
real and imaginary parts of the molar susceptibility are plotted
in the main graph, and the inverse of the real part of the molar
susceptibility is shown in the inset.  The signal in the imaginary
part below $T$ = 2.5\,K indicates dissipation accompanying the
transition.}
\label{fig:Pr6O11PowderSus}
\end{center}
\end{figure}

We measured the magnetic susceptibility of PrO$_2$ using a
commercial SQUID magnetometer, with a powder sample of mass
270\,mg. The measurements were made using the reciprocating sample
option (RSO), with an applied field of $H = 1$\,T.  Data were
taken while cooling in steps from $T = 350$\,K to $T = 2$\,K, with
a delay to allow temperature equilibriation at each step.  A plot
of the molar susceptibility (per mole Pr) is shown in Fig.\
\ref{fig:PrO2PowderSus}. The inset shows the inverse molar
susceptibility.  Three features are evident.  First, there is a
peak at $T = 14$\,K, due to antiferromagnetic ordering.  Second,
there is a change in gradient at $T = 122 \pm 2$\,K which is seen
most clearly in the inverse susceptibility.  Following our
structural investigation we can identify this with the internal
distortion at $T_{\rm D}$.  This anomaly had not been noticed in
previous measurements, \cite{Kern:1964, MacChesney} due to the
intrinsic noise present. Third, there is an upturn in the
susceptibility below $T \approx 6$\,K which is due to the presence
of a small amount of Pr$_6$O$_{11}$ in the sample. We demonstrate
this in Figure \ref{fig:Pr6O11PowderSus}, which shows a plot of
the magnetic susceptibility of Pr$_6$O$_{11}$ measured by AC
susceptometry with a powder sample of mass 327\,mg. Between 1.5\,K
and 2.5\,K there is evidence for a magnetic transition with glassy
characteristics. At 2\,K the molar susceptibility (per mole Pr) of
Pr$_6$O$_{11}$ is $\sim 100$ times that of PrO$_2$, so the upturn
below 6\,K in the PrO$_2$ susceptibility is consistent with the
1\% Pr$_6$O$_{11}$ impurity known to be present in the sample.
There is no feature in the Pr$_6$O$_{11}$ susceptibility at
$T_{\rm D}$.

We used our measurement of the PrO$_2$ susceptibility to calculate
the effective paramagnetic moment of the Pr ion at various
temperatures, assuming that the susceptibility obeys the
Curie-Weiss law for an antiferromagnet above $T_{\rm N}$:

\begin{equation}
\chi^{\rm mol} = \frac{\mu_{\it eff}^2N_{\rm A}}{30k_{\rm
B}(T+\theta)},
\label{eq:Curie}
\end{equation}

\noindent where $\chi^{\rm mol}$ is the molar magnetic
susceptibility, $\theta$ is the Weiss constant, $N_{\rm A}$ is
Avogadro's number, $k_{\rm B}$ is Boltzmann's constant and
$\mu_{\it eff}$ is the effective paramagnetic moment of the Pr
ion.  $\chi^{\rm mol}$ is in cgs units (emu\,[mol Pr]$^{-1}$), as
dictated by convention, whereas the quantities on the right hand
side of Equation \eqref{eq:Curie} are in SI units. The conversion
factor from SI to cgs units is incorporated into the right hand
side.

Over the temperature range $T = 250$--350\,K we obtained $\mu_{\it
eff} = 2.32 \mu_{\rm B}$.  This is close to the value of $2.54
\mu_{\rm B}$ expected for a free Pr$^{4+}$ ion.  Below $T_{\rm D}$
the gradient of the inverse susceptibility first decreases then
increases, causing $\mu_{\it eff}$ to decrease.  This indicates a
reduction in the number of magnetic degrees of freedom, which is
probably caused by a lifting of the degeneracy of the crystal
field ground state by the structural distortion (it is generally
accepted that the ground state is a $\Gamma_8$ quartet in the
fluorite phase). \cite{Kern:1984, Boothroyd:2001}

Although an anomaly at $T = 122$\,K had not been noticed in
previous measurements, the curvature of the susceptibility trace
below this temperature had been noticed previously by Kern.
\cite{Kern:1964}  When attempting a crystal field analysis he
found that his data could not be fitted well with a crystal field
of cubic symmetry.  He proposed that the oxygen ions surrounding
the Pr ion did not in fact form a cube, but that the two at
opposite ends of the body diagonal were displaced outwards along
the (111) direction, causing a splitting of the $\Gamma_8$ crystal
field level into two doublets. Using this modified structure he
obtained a good fit to his data over the whole temperature range
0--300\,K.  He calculated the splitting of the $\Gamma_8$ level to
be 28.8\,meV.

The significance of this result was later dismissed
\cite{MacChesney} because the sample used did not exhibit a clear
antiferromagnetic transition.  Although a discontinuity was
observed at 14\,K, the susceptibility continued to rise at lower
temperatures.  The sample had been prepared from a starting
material of Pr$_6$O$_{11}$ by annealing at a temperature of
360$^{\circ}$C under 5 atmospheres of oxygen pressure.  These
conditions lie close to the border, on a temperature-pressure
diagram, of the regions in which the PrO$_2$ and Pr$_6$O$_{11}$
phases are stable, so it is likely that the sample was not single
phase PrO$_2$.  However, x-ray diffraction showed that it had a
lattice parameter of $5.393 \pm 0.007$\,\AA, which agrees well
with the accepted value for PrO$_2$ of $5.393 \pm 0.001$\,\AA, and
is much less than the value for Pr$_6$O$_{11}$ of 5.468\,\AA.
\cite{Eyring}  This suggests that the sample was mostly PrO$_2$,
but contained enough of the Pr$_6$O$_{11}$ phase to obscure
partially the antiferromagnetic transition.  By adding proportions
of our susceptibility data for Pr$_6$O$_{11}$ and PrO$_2$, and
comparing the results with the data obtained by Kern, we estimate
that his sample contained $\sim$ 20\% Pr$_6$O$_{11}$.

Since we believe that Kern's sample was predominantly PrO$_2$, and
we have now obtained direct evidence for an internal distortion of
the oxygen sublattice below $T_{\rm D} = 120$\,K, we believe that
his original crystal field analysis, in particular the 28.8\,meV
splitting of the $\Gamma_8$ crystal field ground state is still of
some relevance below $T_{\rm D}$.

\subsection{Specific heat capacity}

We measured the specific heat capacity of PrO$_2$ by the
relaxation method \cite{Bachmann} using a laboratory-built
calorimeter. The sample was a disc-shaped pressed pellet of
PrO$_2$ powder of diameter 10\,mm, thickness 1\,mm and mass
83\,mg, which was mounted on the sapphire sample platform of the
calorimeter using a small amount of Apiezon grease (high thermal
conductivity grease).

The measurements were made over a temperature range 2.4\,K to
23\,K.  We measured the heat capacity of the sapphire platform
first, then the heat capacity of the sample and platform combined.
We obtained the heat capacity of the sample alone by subtracting
the former from the latter.  Finally, we obtained the specific
heat capacity of the sample by dividing by the sample mass.

\begin{figure}[!ht]
\begin{center}
\includegraphics{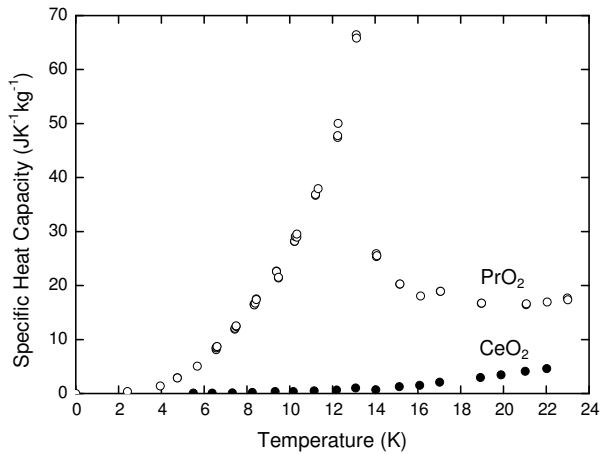}
\caption{Specific heat capacity of PrO$_2$ and CeO$_2$.  The white
circles are the PrO$_2$ data and the black circles are the CeO$_2$
data. The CeO$_2$ data provide a good estimate of the phonon
contribution to the PrO$_2$ specific heat.}
\label{fig:PrO2SpecHeat}
\end{center}
\end{figure}

Figure \ref{fig:PrO2SpecHeat} shows a plot of the specific heat
capacity of PrO$_2$ versus temperature.  The experimental
uncertainty is estimated to be $\sim$ 5\%.  A lambda anomaly due
to the antiferromagnetic ordering is observed at $\sim 13.5$\,K.
This is superimposed on the contribution to the specific heat from
vibrations of the crystal lattice.

The specific heat capacity of a pressed pellet of CeO$_2$ powder
(commercially obtained) was also measured.  This sample had mass
81\,mg and was of similar shape to the PrO$_2$ sample.  Since
CeO$_2$ is non-magnetic, but has the same crystal structure, a
similar lattice parameter and similar formula mass to PrO$_2$, it
provides a good estimate of the contribution of the crystal
lattice to the PrO$_2$ specific heat.  The CeO$_2$ data is plotted
with the PrO$_2$ data in Fig.\ \ref{fig:PrO2SpecHeat}.

To determine the degeneracy of the crystal field ground state in
PrO$_2$, we used our measurement of the specific heat capacity to
calculate the change in magnetic entropy on passing through the
magnetic transition.  The specific entropy is related to the
specific heat capacity by the formula

\begin{equation}
s(T) = \int_0^T \frac{c}{T}\,\dif T,
\end{equation}

\noindent where $c$ is the specific heat and $s$ is the change in
specific entropy between $T$ = 0 and a finite temperature $T$. The
total specific entropy of PrO$_2$ (magnetic entropy $+$ lattice
contribution) was obtained from the specific heat data by dividing
by $T$, then integrating by the extended trapezoidal rule.
\cite{NumericalRecipes} This gave

\begin{equation}
\Delta s_{\rm total} = \int_0^{23} \frac{c({\rm PrO_2})}{T}\,\dif
T =\;32.9\,\rm{JK^{-1}kg^{-1}}.
\label{eq:TotalSpecHeat}
\end{equation}

\noindent To obtain the change in magnetic entropy it was
necessary to subtract off the lattice contribution.  This was
estimated using the CeO$_2$ data:

\begin{equation}
\Delta s_{\rm lattice} = \int_0^{23} \frac{c({\rm
CeO_2})}{T}\,\dif T =\;1.7\,\rm{JK^{-1}kg^{-1}}.
\label{eq:PhononSpecHeat}
\end{equation}

\noindent By subtracting \eqref{eq:PhononSpecHeat} from
\eqref{eq:TotalSpecHeat} we obtained the change in magnetic
entropy:

\begin{equation}
\Delta s_{\rm magnetic} = \Delta s_{\rm total} - \Delta s_{\rm
lattice} =\;31.2\,\rm{JK^{-1}kg^{-1}}.
\label{eq:change}
\end{equation}

\noindent From Boltzmann's law,

\begin{equation}
\Delta s_{\rm magnetic} = nk_{\rm B}\ln g,
\end{equation}

\noindent where $n$ is the number of magnetic ions per unit mass
(for PrO$_2$, $n = 3.48 \times 10^{24}$\,kg$^{-1}$) and $g$ is the
degeneracy of the ground state above $T_{\rm N}$ (below $T_{\rm
N}$ the degeneracy is lifted). The entropy changes predicted for a
quartet ($g = 4$) and a doublet ($g = 2$) ground state are
66.7\,JK$^{-1}$kg$^{-1}$ and 33.3\,JK$^{-1}$kg$^{-1}$
respectively.   The measured change \eqref{eq:change} suggests
that the ground state is a doublet.  This finding supports a
distortion of the crystal structure, which would lift the
degeneracy of the $\Gamma_8$ quartet below $T_{\rm D}$.  An
extension of the specific heat capacity measurement up to room
temperature, to look for an anomaly near 120\,K, would be
eminently worthwhile.

\subsection{Electrical conductivity}

We measured the electrical conductivity of PrO$_2$ as a function
of temperature using a simple probe designed for use with a helium
cryostat.  Two thin gold wires were attached to opposite sides of
the single crystal sample using silver conducting paint (attempts
to attach four wires to the tiny crystal were unsuccessful, due to
the difficulty in achieving good electrical contact while keeping
the contacts separate).  The crystal was glued to the copper base
of the probe with GE varnish to ensure good thermal contact.

\begin{figure}[!ht]
\begin{center}
\includegraphics{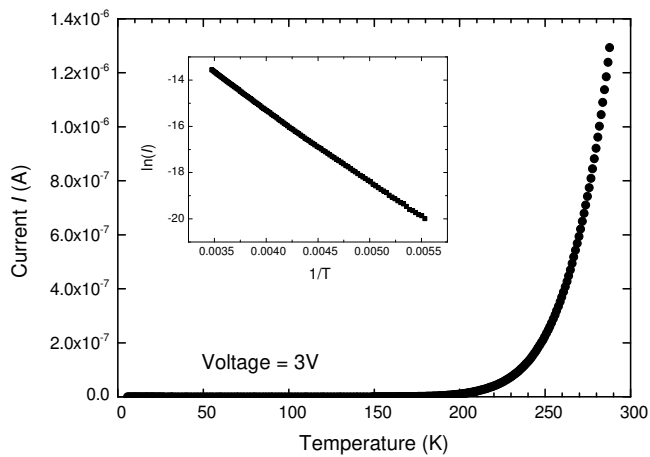}
\caption{Current $I$ through a single crystal of PrO$_2$ at a
constant voltage of 3\,V.  The conductivity is proportional to
$I$. The inset shows $\ln I$ against $1/T$.}
\label{fig:Conductivity}
\end{center}
\end{figure}

Four-terminal measurements were made of the resistance of the
sample + gold wires (this was justified, since the resistance of
the crystal was much greater than the resistance of the gold
wires) over a temperature range from 5.8\,K to 288\,K. The
temperature was increased in steps, allowing the sample to
equilibriate at each new setpoint.  Figure \ref{fig:Conductivity}
shows a plot of current against temperature at a voltage of 3\,V.
Below $T \sim$ 180\,K the current became too small to measure
reliably. We were unable to determine the absolute value of the
electrical conductivity because the shape of the crystal was
irregular and its dimensions were not known exactly.

The inset to Fig.\ \ref{fig:Conductivity} shows that the
conductivity has an activated behaviour, at least over the
temperature range 180--288\,K.  To determine the activation energy
we fitted our data using the Arrhenius form:

\begin{equation}
I = I_0\exp\left(-\frac{E_{\rm a}}{k_{\rm B}T}\right),
\end{equation}

\noindent where $I$ is the current and $E_{\rm a}$ is the
activation energy.  We extracted $E_{\rm a}$ from several runs,
each measured at a different constant voltage (between $-5$\,V and
$+5$\,V), and took the mean of these values to obtain $E_{\rm a}$
= 0.262\,$\pm$\,0.003\,eV for the temperature range 180--288\,K.
This is much smaller than the band gap of $\sim 2.5$\,eV predicted
by band structure calculations, \cite{Koelling} and is also
smaller than the activation energies of other PrO$_x$ materials,
which have been reported in the range 0.4--1\,eV.
\cite{Eyring:1979, SubbaRao, Arakawa}

Possible mechanisms for electrical conductivity in PrO$_2$ are (i)
ionic conductivity, (ii) intrinsic electronic conductivity due to
hopping between Pr sites, (iii) extrinsic electronic conductivity
due to impurities.  Compounds with the fluorite structure often
exhibit ionic conduction when vacancies caused by Schottky or
Frenkel defects allow the oxygen ions to hop from site to site.
However, at room temperature the number of thermally induced
defects is very small, so ionic conductivity in PrO$_2$ is
unlikely to be important in this temperature range.  It has been
suggested that ionic conductivity in PrO$_x$ materials is
significant only above $\sim 900$\,K. \cite{SubbaRao, Arakawa}
Therefore, it is most likely that the conductivity in PrO$_2$ is
electronic.  Given the small activation energy found here, the
mechanism warrants further investigation.

\section{Discussion and conclusions}

We have presented neutron diffraction experiments on single
crystal PrO$_2$, which reveal an internal distortion of the oxygen
sublattice at $T_{\rm D} = 120 \pm 2$\,K and a related component
of the magnetic structure below $T_{\rm N} = 13.5 \pm 0.2$\,K. The
displacements of the oxygen ions are five times as large as those
observed in UO$_2$, \cite{Faber} and this is reflected in the high
value of the transition temperature.  The observation that both
the distorted crystallographic structure and the related component
of the magnetic structure have a unit cell that is doubled along
one crystal axis with respect to the fluorite structure suggests
that the displacement of the oxygen ions affects the magnetic
ordering of the Pr sublattice.

Our measurement of the specific heat capacity indicates that the
crystal field ground state of the Pr 4$f$ electron is a doublet,
consistent with the lowering of the local symmetry of the Pr site.
This raises the question of what mechanism drives the structural
distortion. The most obvious answer is that the electronic energy
is reduced by a collective Jahn-Teller distortion at the expense
of a small penalty in elastic energy. Another possibility is that
the distortion is a consequence of a quadrupolar ordering of the
Pr 4$f$ orbitals. Intuitively this would seem an unlikely
mechanism, since the coupling between electric quadrupoles would
be expected to be too weak to cause an orbital ordering at a
temperature as high as 120\,K. Nevertheless, there is evidence for
a degree of Pr 4$f$---O 2$p$ hybridisation in PrO$_2$,
\cite{Karnatak, Bianconi} and such a covalency effect could
provide a mechanism for aligning the 4$f$ orbitals.

The splitting of the ground state could also partly explain the
observation of an ordered magnetic moment smaller than that
associated with the $\Gamma_8$ quartet ground state of cubic
symmetry. On the other hand, we have previously reported a broad
continuum of magnetic scattering in the excitation spectrum of
PrO$_2$, which we believe arises from a strong magnetoelastic
coupling. \cite{Boothroyd:2001} This coupling would tend to quench
magnetic degrees of freedom and reduce the size of the ordered
moment by the dynamic Jahn-Teller mechanism. From the experimental
evidence available so far it seems likely that both the static and
dynamic Jahn-Teller effects are important in PrO$_2$, but it is
not clear which of the two is dominant at low temperatures.

In work that we will report elsewhere,
\cite{Gardiner:PrO2MagneticField} we have found that the
application of a magnetic field in the distorted phase at $T <
T_{\rm D}$ produces striking hysteresis effects in both the
magnetic structure and the magnetisation. The totality of results
that have emerged in recent years suggests that a strong interplay
between electronic and lattice degrees of freedom exists in
PrO$_2$ and influences its properties, perhaps in unexpected ways.

\begin{acknowledgments}
We would like to thank P. Santini for insightful discussions, and
D. Prabhakaran and F. Wondre for help with sample preparation and
characterisation.  We would also like to thank J. Jensen for
suggesting the chiral model for the distorted crystal structure
and making many other helpful comments.  We are also grateful to
M. Enderle and K. Kiefer for advice on the design of the
calorimeter.  Financial support and provision of a studentship for
CHG by the EPSRC is acknowledged.
\end{acknowledgments}

\bibliography{PrO2PrBCO}

\end{document}